\documentclass[prx,showpacs,twocolumn,amsmath,amssymb,floatfix]{revtex4-1}

\usepackage{graphicx}
\DeclareGraphicsExtensions{.png,.jpg,.eps,.pdf}
\usepackage{bm}
\usepackage{color}
\usepackage[applemac]{inputenc}
\newcommand{\New}[1]{#1}

\begin{document} 
   
\title{Majorana Zero Modes in Graphene}

\author{P. San-Jose$^1$, J.  L. Lado$^2$, R. Aguado$^1$, F. Guinea$^{3,4}$, J. Fern\'andez-Rossier$^{2,5}$}
\affiliation{
$^1$Instituto de Ciencia de Materiales de Madrid, Consejo Superior de Investigaciones Cient\'ificas (ICMM-CSIC), Sor Juana In\'es de la Cruz 3, 28049 Madrid, Spain\\
$^2$International Iberian Nanotechnology Laboratory (INL), Av. Mestre Jos\'e Veiga, 4715-330 Braga, Portugal\\
$^3$Instituto Madrile\~no de Estudios Avanzados en Nanociencia (IMDEA-Nanociencia), 28049 Madrid, Spain\\
$^4$Department of Physics and Astronomy, University of Manchester, Manchester M13 9PL, UK\\
$^5$Departamento de Fisica Aplicada, Universidad de Alicante, 03690 Alicante, Spain}

\begin{abstract}
A clear demonstration of topological superconductivity (TS) and Majorana zero modes remains one of the major pending goal in the field of topological materials. One common strategy to generate TS is through the coupling of an s-wave superconductor to a helical half-metallic system. Numerous proposals for the latter have been put forward in the literature, most of them based on semiconductors or topological insulators with strong spin-orbit coupling. Here we demonstrate an alternative approach for the creation of TS in graphene/superconductor junctions without the need of spin-orbit coupling. Our prediction stems from the helicity of graphene's zero Landau level edge states in the presence of interactions, and on the possibility, experimentally demonstrated, to tune their magnetic properties with in-plane magnetic fields. We show how canted antiferromagnetic ordering in the graphene bulk close to neutrality induces TS along the junction, and gives rise to isolated, topologically protected Majorana bound states at either end. We also discuss possible strategies to detect their presence in graphene Josephson junctions through Fraunhofer pattern anomalies and Andreev spectroscopy. \New{The latter in particular exhibits strong unambiguous signatures of the presence of the Majorana states in the form of universal zero bias anomalies.}  Remarkable progress has recently been reported in the fabrication of the proposed type of junctions, which offers a promising outlook for Majorana physics in graphene systems.
\end{abstract}

\date{\today} 

\maketitle

\section{Introduction}

The realisation of topological superconductivity (TS), a novel electronic phase characterised by Majorana excitations, has become a major goal in modern condensed matter research. Despite promising experimental progress\cite{Mourik:S12,Deng:NL12,Das:NP12,Rokhinson:NP12,Finck:PRL13,Churchill:PRB13,Lee:NN14,Hart:NP14,Pribiag:NN15,Nadj-Perge:S14,Pawlak:15,Kurter:NC15,Wiedenmann:15} on a number of appealing implementations\cite{Fu:PRL08,Sau:PRL10,Lutchyn:PRL10,Oreg:PRL10}, a conclusive proof of  TS remains an open challenge. We here report on a new approach to obtain TS and Majorana states in graphene/superconductor junctions. Key to our proposal is the interaction-induced magnetic ordering of graphene's zero Landau level (ZLL). Coupling this unique state to a conventional superconductor gives rise to novel edge states whose properties depend on the type of magnetic order. In particular, the canted antiferromagnetic phase is a natural host for Majorana bound states. Our proposal combines effects that were recently demonstrated experimentally (tunable spin ordering of the ZLL \cite{Young:NP12,Young:N14} and ballistic \cite{Calado:15,Shalom:15} graphene/superconductor junctions of high-transparency\cite{Shalom:15} operating in the Quantum Hall regime \cite{Calado:15}), and is thus ready to be tested.

While intrinsic TS is rare, it can be synthesised effectively through the coupling of a conventional s-wave superconductor (SC) and tailored electronic gases with spin-momentum locking. Using this recipe, it has been predicted that Majorana excitations should emerge when one induces superconductivity onto topological insulators \cite{Fu:PRL08} or semiconductors with strong spin-orbit coupling \cite{Sau:PRL10}. Particularly attractive are implementations of one-dimensional TS using either semiconducting nanowires \cite{Lutchyn:PRL10,Oreg:PRL10} or edge states in two-dimensional Quantum Spin Hall (QSH) insulators, since the main ingredients are already in hand. These ideas have spurred a great deal of experimental activity \cite{Mourik:S12,Deng:NL12,Das:NP12,Rokhinson:NP12,Finck:PRL13,Churchill:PRB13,Lee:NN14,Hart:NP14,Pribiag:NN15,Nadj-Perge:S14,Pawlak:15,Kurter:NC15,Wiedenmann:15}. Despite this progress, however, an unambiguous demonstration of TS is, arguably, still missing. Important limitations of these systems include disorder, bulk leakage, or imperfect proximity effect (the so-called soft gap problem). Thus, it is worthwhile to explore alternative materials. 

One particularly interesting option is graphene \cite{Neto:RMP09}, which exhibits very large mobilities even in ambient conditions, and where a ballistic proximity effect has been recently demonstrated \cite{Shalom:15}. Graphene was the first material where a topological insulating phase was proposed \cite{Kane:PRL05} in the presence of a finite intrinsic spin-orbit coupling. Kane and Mele showed that graphene then becomes gapped around neutrality, and a single helical edge mode with spin locked to propagation direction develops at each edge. In such QSH regime, gapping the edge states through proximity to a conventional superconductor gives rise to a one-dimensional TS along the interface \cite{Fu:PRL08}. Graphene's negligible spin-orbit coupling, however, has proved to be a fundamental roadblock in this programme.

\begin{figure*}
\includegraphics[width=\textwidth]{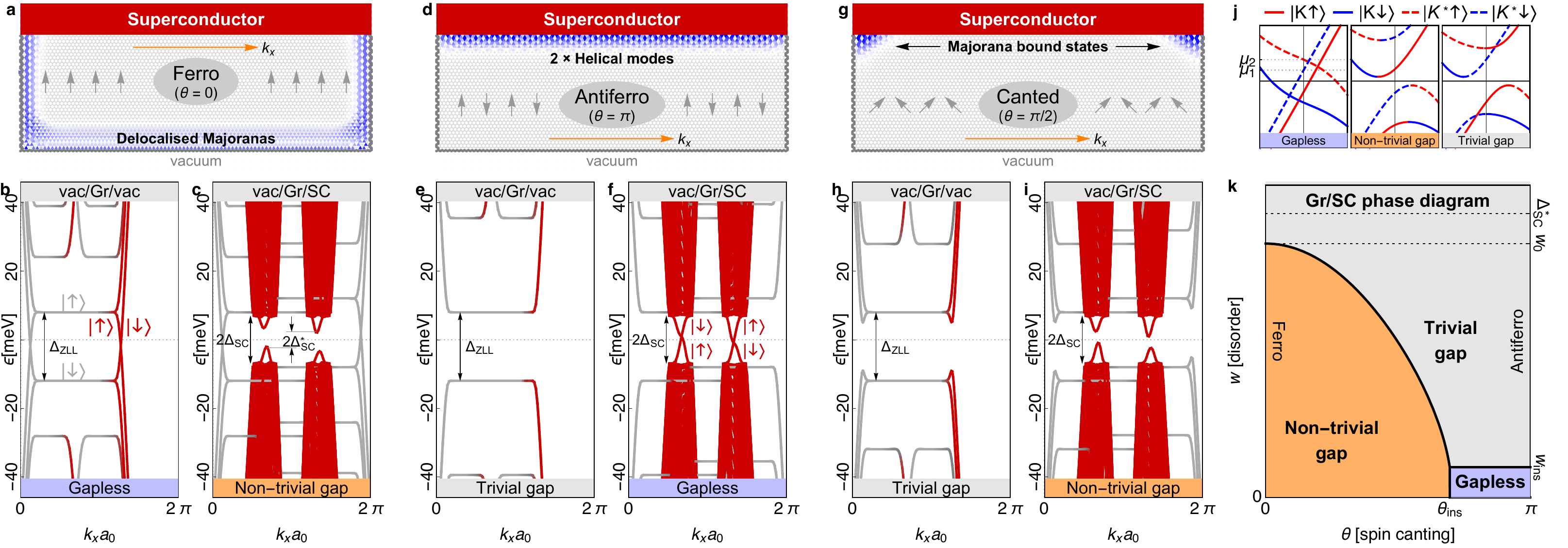}
\caption{
Sketch, bandstructure and phase diagram of a \New{350 nm-wide} graphene sample \New{(Fermi velocity $v_F=10^6\,m/s$)} in the Quantum Hall regime \New{(out of plane field $B_z=1$ T)} in various configurations. The chemical potential (dotted line) is tuned within the gap \New{$\Delta_\mathrm{ZLL}\approx 20 \textrm{meV}$} of the zero Landau level, which has ferromagnetic (a-c), antiferromagnetic (d-f) or canted antiferromagnetic ordering (g-i) due to electronic interactions. The bandstructures under each sketch correspond to an infinite graphene ribbon surrounded by vacuum (vac/Gr/vac panels) or coupled to a superconductor of gap $\Delta_\mathrm{SC}$ \New{(chosen large $\sim 7$ meV for visibility)} along the top edge as in the sketches (vac/Gr/SC panels, Nambu bands). Bands in red correspond to eigenstates localised at the top edge of the ribbon. The zero-energy local density of states is shown in blue in each sketch. (j) Low energy bandstructure of states along a graphene/superconductor interface folded onto the $\Gamma$ point, in the gapless (left panel, zoom of panel f), non-trivially gapped (middle panel, canted order) and trivially gapped (right panel, with strong intervalley scattering) phases.  (k) Phase diagram of said interface, computed from its low energy effective Hamiltonian (see text), as a function of magnetic angle $\theta$ and intervalley coupling $w$. The three possible phases are shown, bounded by threshold values $w_0$, $w_\mathrm{ins}$ and $\theta_\mathrm{ins}$, see main text.}
\label{fig:1}
\end{figure*}

In this work we present a simple mechanism to realise the above situation in graphene without recourse to spin-orbit coupling. We consider a graphene ribbon in the Quantum Hall (QH) regime, in which, unlike in the QSH case, time-reversal symmetry is broken by a strong magnetic flux (Appendix \ref{ap:modelling}). In contrast to conventional two-dimensional electron gases, graphene develops a zero-Landau level at the Dirac point, which has been shown in pristine samples to become split due to electronic interactions \New{\cite{Zhang:PRL06,Bolotin:N09,Miller:S09,Feldman:S12,Kou:S14,Yu:NP14}}. Experimental evidence \cite{Young:N14} points towards spontaneous antiferromagnetic ordering \cite{Herbut:PRB07a,Lado:PRB14}, although other broken symmetries have been discussed \cite{Young:NP12}. In this work we consider all possible magnetic orders. Fig. \ref{fig:1} summarises the different possibilities, ranging from ferromagnetic (F) to antiferromagnetic (AF) ordering \footnote{Recently, schemes to experimentally probe the AF phase through spin excitations have been proposed \cite{Takei:15}.}, including canted AF which may be controlled by an external in plane Zeeman field as argued in Ref. \onlinecite{Young:N14}. The different orders are parameterised by the angle $\theta$ between the spin orientation of the ZLL in the two graphene sublattices, so that $\theta=0$ for F and $\theta=\pi$ for AF. 

\New{While $\theta$ is considered in our model as an externally tuneable parameter as in Ref. \onlinecite{Young:N14}, we have checked that a mean-field calculation in a honeycomb Hubbard model under an in-plane Zeeman field (Appendix \ref{ap:modelling}) yields the same bulk and edge phenomenology presented in this work \cite{Lado:PRB14}. Corrections beyond mean field and the Hubbard model have been explored theoretically in the past, and have predicted the formation of a Luttinger liquid domain on an infinite vacuum edge \cite{Fertig:PRL06}. The corresponding excitation density resembles the non-interacting edge for F order, rather than the AF case. The interacting problem at a highly transparent superconducting contact remains an open problem. We conjecture that, given their robust topological origin, the Majorana phenomena described here at a mean field level would survive in the Luttinger regime, at least within a limited range of parameters, and with power-law corrections to the transport results. These issues, however, remain beyond the scope of this work.} 

\section{Topological superconductivity in Quantum Hall graphene}

An infinite ferromagnetically ordered ($\theta=0$) ribbon in vacuum 
has a QH mean-field bandstructure \cite{Abanin:PRL06} as shown in Fig. \ref{fig:1}b. \New{(Details on the modelling are given in Appendix \ref{ap:modelling}).} The ZLL is spin-split into the two spin sectors in the direction of the ferromagnetic order, denoted by $|\!\!\uparrow\rangle,|\!\!\downarrow\rangle$ at energies $\pm\Delta_\mathrm{ZLL}/2$ respect to the Dirac point. For energies within this gap, a single pair of gapless counterpropagating spin-polarised edge states develop, shown in red for states at the upper vacuum edge of the sample. Note the peculiar situation created in this energy window: edge states are not chiral like in the conventional QH regime, but may rather propagate in both directions with opposite spins, like in the QSH regime. Also, valley degeneracy is lifted at any given edge, which hosts a single state per propagating direction. Upon contacting one edge to a conventional superconductor of gap $\Delta_\mathrm{SC}$ (as in the sketch of Fig. \ref{fig:1}a), while keeping the chemical potential (dotted line) within the ZLL gap, the edge states along the interface \New{develop an induced gap $\Delta_\mathrm{SC}^*$ (see the corresponding Nambu bandstructure of Fig. \ref{fig:1}c -- red lines, once more, indicate states localised at the upper edge of the graphene ribbon, including now the dense quasiparticle spectrum of the superconductor). The gap $\Delta_\mathrm{SC}^*$ is an important scale in this problem, since it turns out to be a topologically non-trivial gap. This is confirmed by computing the bandstructure's $\mathbb{Z}_2$ topological invariant, Eq. (\ref{Pfaffian}), relevant for quasi-one dimensional $D$-class systems \cite{Qi:RMP11}. Unlike in a conventional QSH system, where time-reversal symmetry is required, the vacuum edge states do not immediately develop a topological gap when contacted to the supercoductor, but requires a reasonably good contact instead. On the other hand, while the conventional QSH metal becomes destroyed by any time-reversal-breaking perturbation (such as inelastic scattering or magnetic impurities), which in turn spoil the non-trivial superconducting gap, this is not the case of the the present implementation, which has a broken time-reversal symmetry from the start. Moreover, we emphasize once more that no spin-orbit coupling at all is necessary for $\Delta_\mathrm{SC}^*$ to develop.}


The immediate consequence of a non-trivial gap topology is the appearance of zero energy Majorana bound states (MBSs) at an interface with a trivial insulator, following the bulk-boundary correspondence principle. In this case, however, both ends of the topologically gapped superconducting interface are coupled to a \emph{gapless} vacuum edge, so that the zero modes become delocalised into the continuum away from the interface (see the zero energy local density of states [LDOS] in blue in Fig. \ref{fig:1}a). This situation is similar to the fate of MBSs at the ends of a topological proximised semiconductor nanowires when strongly coupled to a metallic environment. 
\cite{Wimmer:NJP11}

The electronic structure associated to an antiferromagnetic ribbon ($\theta=\pi$, Fig. \ref{fig:1}d) is the opposite. The states along a vacuum edge are now (trivially) gapped \cite{Lado:PRB14} like the ZLL itself (Fig. \ref{fig:1}e). Surprisingly, when contacting the edge to a conventional superconductor \emph{two} pairs of gapless helical edge modes emerge with spin-momentum locking around conjugate momenta $K$ and $K^*$ (Fig. \ref{fig:1}f). These unexpected states, spatially spread along the interface (see the blue LDOS in Fig. \ref{fig:1}d), are decoupled electron-hole ($e$-$h$) superpositions with orthogonal and well defined spin orientation along the AF axis, $|K^{(*)}\!\!\uparrow\rangle=a|\phi_{e\uparrow}\rangle + b|\phi_{h\downarrow}\rangle$ and $|K^{(*)}\!\!\downarrow\rangle=a'|\phi_{e\downarrow}\rangle + b'|\phi_{h\uparrow}\rangle$. A full discussion of these states is presented in Appendix \ref{ap:wavefunction}. The two helical edge modes remain gapless as long as no AF canting is present in graphene ($\theta=\pi$) and intervalley scattering is zero at the interface. We next consider deviations from these two assumptions.

Canting of the AF order may be induced by means of a large enough in-plane Zeeman field, and is thus to some extent externally tunable. This idea was employed in Ref. \onlinecite{Young:N14} to tune a graphene QH bar in vacuum between the AF and F regimes, leading to a insulator-to-helical metal transition in edge transport (evolution from Figs. \ref{fig:1}e to \ref{fig:1}b). A typical canted AF bandstructure ($\theta=\pi/2$) is shown in Fig. \ref{fig:1}h. The vacuum edge states exhibit a topologically trivial and $\theta$-dependent gap, smaller than the bulk $\Delta_\mathrm{ZLL}$. Along a superconductor interface, the canted AF helical states are also gapped, Fig. \ref{fig:1}i. Like in the ferromagnetic case, this gap is topologically non-trivial. This situation allows for the emergence of true localised zero-energy MBSs at the ends of the superconductor interface, where the edge gap changes topology, see Fig. \ref{fig:1}g. The MBSs are topologically protected, and are not destroyed by any small perturbations, or even by modifying the crystal structure of the superconductor (Appendix \ref{ap:square}).

To understand the full phase diagram of the graphene/superconductor interface  quantitatively, it is useful to employ a simplified description in terms of an effective low-energy Hamiltonian for the edge states (see Appendix \ref{ap:Heff}).  The model is valid for a chemical potential tuned to the ZLL gap, and has the advantage of allowing us to incorporate the effects of atomic disorder along the junction (encoded in an intervalley coupling $w$, where `valley' here refers to the conjugate $K$ and $K^*$ momenta) and arbitrary spin canting (encoded in an intravalley splitting $b_\theta=\Delta_\mathrm{SC}^*\cos[\theta/2]$). It correctly describes the three possible phases for low-energy interface modes: gapless, trivially gapped and non-trivially gapped. The corresponding phase diagram is shown in Fig. \ref{fig:1}k. Typical edge-mode dispersions within each phase (with $K$ and $K^*$ points folded onto the $\Gamma$ point) are shown in Fig. \ref{fig:1}j, and are characterised by their energies $\mu_{1,2}<\Delta_\mathrm{SC}^*$ at $k_x=0$ for $\theta=\pi$ [AF], $w=0$, and their corresponding velocities $v_{1,2}>0$ (left panel).

The gapless interface regime (light blue in Fig. \ref{fig:1}k) is achieved for $|w|<w_\mathrm{ins}\equiv\frac{1}{2}|\mu_1\sqrt{v_2/v_1}-\mu_2\sqrt{v_1/v_2}|$ and $
\theta>\theta_\mathrm{ins}$, where $b_{\theta_\mathrm{ins}}=\Delta_\mathrm{SC}^*\cos(\theta_\mathrm{ins}/2)\equiv\frac{1}{2}|\mu_1\sqrt{v_2/v_1}+\mu_2\sqrt{v_1/v_2}|$. The gapped regimes are characterised by the $\mathbb{Z}_2$ topological invariant of the system, which reads $\nu=\mathrm{sign}\left(w^2+\mu_1\mu_2-b_\theta^2\right)$ (see Appendix \ref{ap:Heff}). A trivially gapped phase $\nu=+1$ is reached for strong intervalley coupling $w$ at the interface, while for intervalley scattering below a threshold $w<\sqrt{b_\theta^2-\mu_1\mu_2}$, the interface is one-dimensional TS with invariant $\nu=-1$. The non-trivially gapped regime is most robust against disorder for F order, for which the threshold $w$ reaches its maximum $w_0=\sqrt{(\Delta_\mathrm{SC}^*)^2-\mu_1\mu_2}$. Note that to achieve a non-trivial TS interface, the intervalley coupling $w$ should therefore never exceed the induced gap $\Delta_\mathrm{SC}^*$ (this is always the case for sufficiently transparent junctions).

\section{Experimental signatures of graphene Majoranas}

\begin{figure}
\includegraphics[width=\columnwidth]{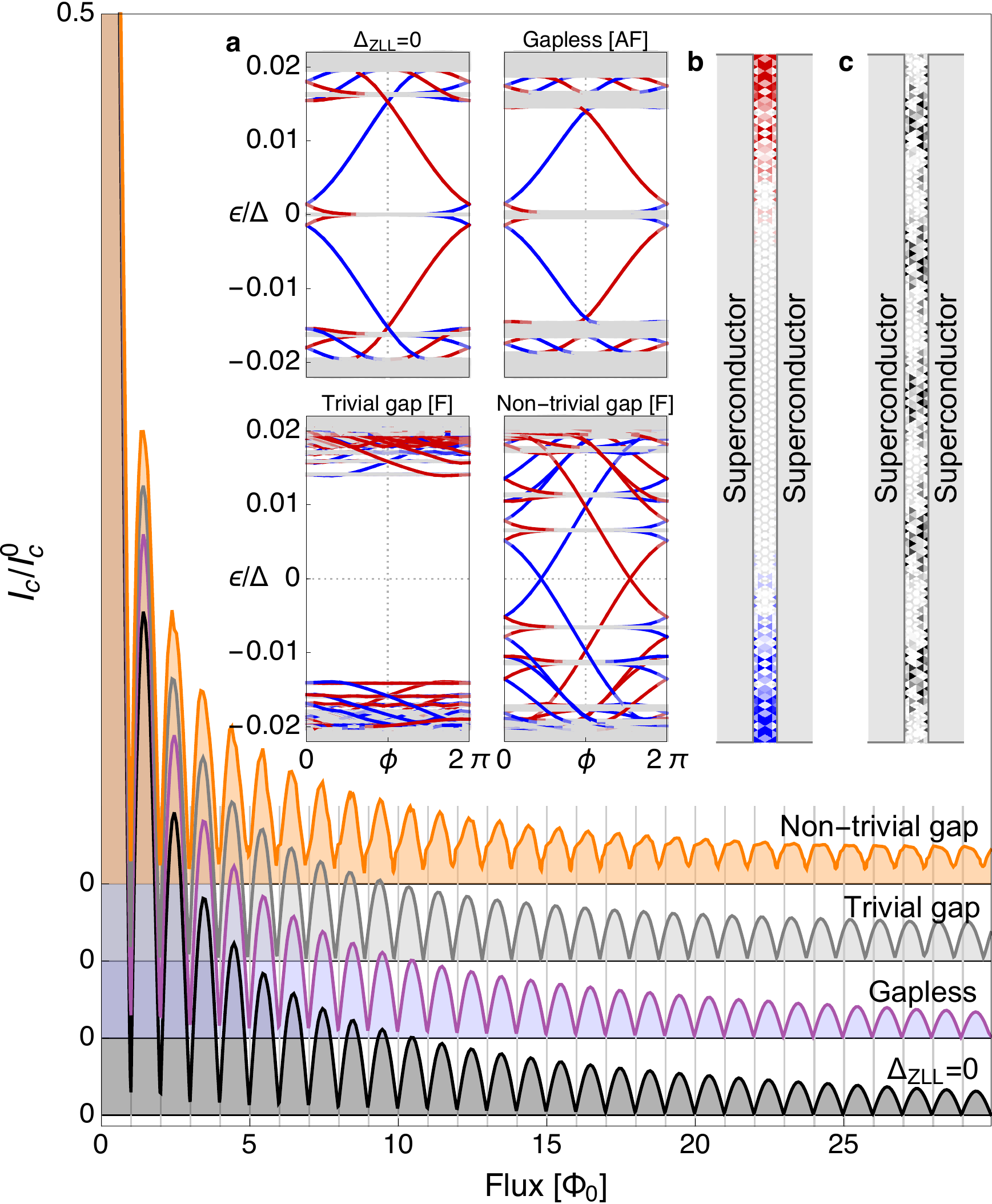}
\caption{
Normalised critical current $I_c/I_c^0$ as a function of magnetic flux through a Josephson junction. Curves from bottom to top (shifted for better visibility) correspond to the non-interacting case (black), gapless AF phase (purple), a trivially gapped AF phase (light gray) and a ferromagnetic phase with a topologically non-trivial gap (orange). Only the latter shows non-decaying non-zero minima in $I_c$, a consequence of an Andreev spectrum (inset $a$) with an odd number (one) of edge-resolved zero energy crossings (bottom-right panel). States coloured in red and blue are located at the top and bottom edges, respectively (inset $b$), while states in gray are spread across the width of the junction (inset $c$).}
\label{fig:2}
\end{figure}

\begin{figure*}
\includegraphics[width=\textwidth]{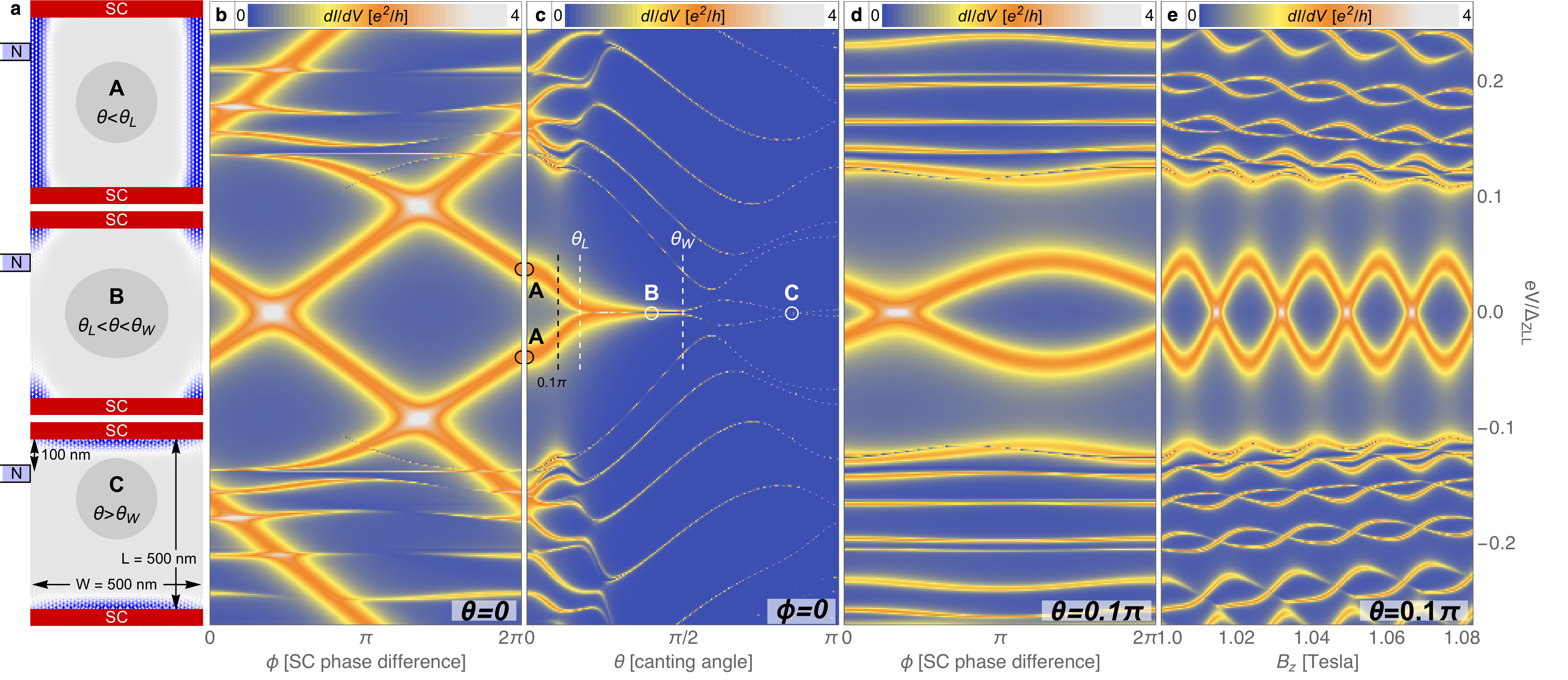}
\caption{\New{(a) In blue, density of the lowest Andreev level formed by the hybridization of four Majorana bound states in a square 500 nm $\times$ 500 nm Josephson junction at different values of canting angle $\theta$. Rest of parameters like in Fig. \ref{fig:1}. In $A$ and $C$ ($\theta<\theta_L$ and $\theta>\theta_W$ respectively) the hybridization is strong, while in $B$ it is exponentially suppressed, and the level remains a four-fold degenerate zero mode. (b-e) Transport spectroscopy $dI/dV$ from a normal point contact $N$ (transparent, single spinful channel), see (a). (b) and (d) show (at $\theta=0$ and $\theta=0.1\pi$, respectively) the $dI/dV$ as a function of junction phase difference $\phi$. Both exhibit an odd number (one) of zero energy crossings of edge-resolved $dI/dV$ resonances, signalling a topologically non-trivial system. (c) $dI/dV$ as a function of canting angle $\theta$ at $\phi=0$. It exhibits a zero bias anomaly in the window $\theta_L<\theta<\theta_W$, see dashed lines. 
(e) The $dI/dV$ dependence with out of plane magnetic field $B_z$ through the sample ($\phi=0, \theta=0.1\pi$)  shows the same zero-crossings as with $\phi$ each time the total flux $\Phi=B_z L W$ is increased by a flux quantum $\Phi_0$.
}}
\label{fig:3}
\end{figure*}

We finally consider measurable signatures of the MBSs in the system. 
A \New{powerful} probe whose feasibility has been recently demonstrated experimentally \cite{Calado:15,Shalom:15} involves \New{interferometry of} critical currents in Josephson junctions, and Fraunhofer pattern anomalies in particular \cite{Allen:15,Kurter:NC15, Pribiag:NN15}. Lee \emph{et al.} predicted \cite{Lee:PRL14} that topological superconductivity in a short and wide Josephson junction could be directly detected in its Fraunhofer pattern, in the form of non-vanishing minima of the critical current  for arbitrary magnetic flux through the junction. Such Fraunhofer anomaly was recently observed in a three-dimensional topological insulator, and was interpreted as possible evidence of MBSs \cite{Kurter:NC15}.

Fig. \ref{fig:2} shows the Fraunhofer pattern in a short and wide ($10\mathrm{nm}\times 3\mathrm{\mu m}$) graphene Josephson junction in various regimes. The black (bottom) curve corresponds to the non-interacting case (no magnetic ordering, $\Delta_\mathrm{ZLL}=0$), which exhibits the conventional $I_c(\Phi)=I_c^0|\sin(\pi\Phi/\Phi_0)|/(\pi\Phi/\Phi_0)$ critical current that decays as the inverse magnetic flux $\Phi$ through the junction $I_c\sim 1/\Phi$ and vanishes at multiples of the flux quantum $\Phi_0=h/2e$. A similar behaviour is obtained in the gapless regime $\theta>\theta_\mathrm{ins}, w<w_\mathrm{ins}$ (purple curve, with $\theta=\pi$ [AF] and $w=0$). In both cases, the junction is host to a narrow quasi-continuum of Andreev bound states around the Fermi energy for any value of the superconducting phase difference $\phi$. The corresponding spectra are shown in the top row of inset $a$ (for $\Phi=15.5\Phi_0$). States in red and blue are localised at the top and bottom edges of the junction (inset $b$), respectively, with gray denoting states spread across the junction (inset $c$).  The case with trivially gapped interfaces (strong intervalley scattering) also exhibits a generic Fraunhofer pattern with vanishing minima (light gray curve, $\theta=0$ [F] and $w>w_0$). The minima, however, occur at fluxes that are shifted away from integer $\Phi/\Phi_0$ at high $\Phi$, while the maxima do not decay like the conventional $I_c\sim 1/\Phi$ pattern, which is connected to non-uniform currents across the junction\cite{Allen:15}. The Andreev spectrum of the trivially gapped phase is qualitatively different from the gapless spectra, and generally shows a distinct gap devoid of any edge states (inset $a$, bottom left). For certain values of parameters, it may exhibit zero-energy crossings inside the gap, but in such cases these crossings are accidental (not topologically protected) and there is always an even number of them at a given edge.

The Josephson junction with a non-trivial gap along the contacts is distinctly different from all previous cases. This phase develops two MBSs at each edge (top and bottom) which hybridise to carry a finite supercurrent that never vanishes as the flux increases. The corresponding Fraunhofer pattern thus exhibits a finite background with a superimposed non-decaying oscillation (orange curve in Fig. \ref{fig:2}), as described in Ref. \onlinecite{Lee:PRL14}.  The finite minima are roughly one half of the maxima at large flux, and occur away from integer $\Phi/\Phi_0$, at values very close to the zeroes of the trivially gapped phase. \New{(Note, however, that such distinctive pattern develops only in junctions shorter than the Majorana localization length and pierced by a large number of flux quanta, so that the Majoranas are well developed and opposite edges are decoupled).} These Fraunhofer anomalies, although not completely unambiguous, thus constitute a measurable hint of the presence of MBSs at the end of a graphene/superconductor interface. \New{Unfortunately, fabricating a very short junction is challenging, in particular due to charge-transfer effects from the superconductors which were neglected here, and which will dope graphene away from neutrality within a few nanometers of the contacts. It is thus important to explore other less stringent experimental schemes that are at the same time not ambiguous. The key is to probe the Andreev spectrum directly for signatures of Majoranas and non-trivial topology.}

\New{The presence of the two hybridised Majoranas per vacuum edge in the non-trivial phase manifests in the Andreev spectrum as a single topologically protected zero energy crossing at each edge as $\phi$ is increased by $2\pi$ (one red and one blue crossing, see bottom-right inset $a$ in Fig. \ref{fig:2}). An odd number of such zero energy crossings has been shown \cite{Kitaev:PU01} to be an direct manifestation of non-trivial topological order $\nu=-1$, and is the underlying reason for the anomalous Fraunhofer pattern of the junction. A completely non-ambiguous demonstration of the presence of MBSs is also thus possible in principle, by directly counting edge-resolved zero-energy crossings using Andreev spectroscopy \cite{Dirks:NP11} in a phase-controlled Josephson junction. This may be achieved by measuring differential conductance $dI/dV$ through a normal point contact attached to one edge of the junction, as sketched in Fig. \ref{fig:3}(a). We assume a single spinful channel is open. Each Andreev level of energy $\epsilon$ in the junction is detected as a $dI/dV$ resonance through the probe at bias $V=\epsilon/e$, with a resonance width that measures the state's probability density at the point contact. Figs. \ref{fig:3}(b,d) show a simulation (see Appendix \ref{ap:dIdV} for details) of such a $dI/dV$ as a function of $\phi$ (at $\theta=0$ and $\theta=0.1\pi$ respectively) for a square 500 nm$\times$ 500 nm Josephson junction. Note that, unlike in the Fraunhofer simulation, this is not a short junction, since that is no longer a desirable or realistic requirement in the context of Andreev spectroscopy. The number of edge-resolved zero energy crossings as $\phi$ is swept from $0$ to $2\pi$ is one, as corresponds to Majorana-hosting SC contacts (compare this $dI/dV (\phi)$ to the blue lines in bottom-right inset Fig. \ref{fig:2}a [short junction]). Incidentally, the $dI/dV (\phi)$ profiles at zero (non-zero) $\theta$, panel b (d), follow the characteristic Andreev level spectra in topological Josephson junctions through long semiconducting nanowires at perfect (non-perfect) transparency \cite{Kwon:EPJB03,Fu:PRB09,San-Jose:PRL12a,San-Jose:PRL14,Cayao:PRB15}. The conductance at the crossings is pinned to a universal value $4e^2/h$ (white spot), see Appendix \ref{ap:dIdV}.} 

\New{While a $\phi$-controlled junction typically requires a SQUID-like geometry and may be experimentally challenging, the existence of MBSs in the junction may be detected even more simply by varying the out of plane magnetic field $B_z$, and hence the total flux $\Phi$ through graphene. An increase of $\Phi$ by $\Phi_0$ is equivalent to increasing $\phi$ by $2\pi$ for large $\Phi/\Phi_0$. This is shown in Fig. \ref{fig:3}e (compare to Fig. \ref{fig:3}d). Moreover, at fixed $B_z$ and $\phi=0$, the MBSs also show up as a zero-bias $dI/dV$ peak (Fig. \ref{fig:3}c) as $\theta$ is tuned -- by the in-plane Zeeman field $B_\parallel$ -- within a range $\theta_L<\theta<\theta_\mathrm{W}\leq\theta_\mathrm{ins}$ (dashed lines). For the realistic parameters used in Fig. \ref{fig:3} ($B_z=1$T, $L=W=500$nm), and using $B_0=10$T as the typical in-plane field $B_\parallel$ for complete Ferro polarization, $\theta_L\approx 0.17\pi$ and $\theta_W\approx 0.5\pi$ are reached for $B_\parallel\approx 9.6$T and $B_\parallel\approx 7.1$T, respectively (see Appendix \ref{ap:dIdV}). Inside this window (point B in Figs. \ref{fig:3}a,c), the  four Majoranas are concentrated at the corners of the junction and do not overlap, as they decay within a distance smaller than both the width $W$ and length $L$ of the junction, and hence appear as a sharp zero-bias resonance in the $dI/dV$ with exponentially small splitting. The resonance has a universal magnitude of $2e^2/h$ at low temperatures, see Appendix \ref{ap:dIdV}. This is analogous to the zero-bias anomalies reported in pioneering Majorana experiments on semiconducting nanowires \cite{Mourik:S12}. In contrast, for $\theta<\theta_L$ (point A, bulk approaching ferro-ordering), Majorana pairs overlap along the vacuum edge and produce a split resonance. Likewise for $\theta>\theta_W$ (e.g. point C, SC contact close or inside  the gapless regime), Majoranas overlap along the SC contact, and develop a (roughly $\phi$-independent) splitting, making the $\theta>\theta_L$ junction strictly trivial. Note also the strong suppression of the width in the $dI/dV$ resonances in this case, due to the exponentially small wavefunction amplitude at the point contact in this geometry (dotted lines overlaid for $\theta>0.6\pi$ to improve visibility). Essentially the same $dI/dV(\theta)$ phenomenology is obtained in a setup with a single superconducting contact (which hosts two Majoranas instead of four), see Appendix \ref{ap:dIdV}.}

\section{Discussion}

Our results show that the spontaneous magnetic ordering of the ZLL in graphene enables the creation of topological superconductivity and Majorana states at an interface with a conventional superconductor, even in the absence of spin-orbit coupling in the system. The key is to tune the Fermi energy in the contact into the ZLL gap, and to achieve a good proximity effect therein. The recently characterised samples of Ref. \onlinecite{Calado:15} are good candidates to realise our proposal. Impressive progress in controlling graphene filling  into proximity gaps has also been reported \cite{Efetov:15}. We furthermore showed that non-vanishing and non-decaying supercurrent minima in the Fraunhofer pattern across a depleted graphene Josephson junction constitute a characteristic signal of topological order and the presence of Majorana bound states in the junction \cite{Lee:PRL14}. Fraunhofer patterns of extraordinary quality have been recently reported in high-transparency ballistic graphene Josephson junctions \cite{Shalom:15}. \New{We predict even stronger observable signatures of non-trivial topology in Andreev transport spectroscopy, both in the form of an odd number of $4e^2/h$ edge-resolved zero-bias crossings versus junction phase difference $\phi$ or out of plane magnetic field $B$, and extended $2e^2/h$ zero bias anomalies versus canting angle $\theta$. These experimental probes and the required device parameters are within reach in top laboratories today.} We thus expect that the possibility of tuning graphene/superconducting interfaces into a topological phase hosting Majorana bound states could be tested soon.

\begin{acknowledgments}

P.S.-J and R. A. are grateful to A. Cortijo and L. Brey for
stimulating discussions. We acknowledge the support of
the European Research Council (F.G.), the Spanish Ministry
of Economy and Innovation through Grants No.
FIS2011-23713 (F.G and P. S.-J) and
FIS2012-33521 (R. A.), the Ram\'on y Cajal programme (P. S.-J), the Marie-Curie-ITN 607904-SPINOGRAPH
(J.L. and J.F.R.),  Generalitat Valenciana (ACOMP/2010/070), and
Framework Programme FP7/2007-2013/ under REA Grant Agreement
No. 607904-13, Prometeo (J.F.R).

\end{acknowledgments}

\appendix


\section{Modelling}
\label{ap:modelling}

In this section we present the system models employed in this work.
A non-interacting graphene flake may be modelled by a nearest-neighbour tight-binding Hamiltonian in an honeycomb lattice, with lattice constant $a_0$
\begin{eqnarray}
H_0&=&-\sum_{\langle i,j\rangle,s}t e^{i\phi_{ij}}c^\dagger_{js}c_{is}+\sum_{is}\mu_N n_{is}
\end{eqnarray}
where $n_{is}=c^\dagger_{is}c_{is}$, $i$ is the site index, $s$ is spin, and $\phi_{ij}=-\frac{e}{\hbar}\int_{\vec r_i}^{\vec r_j}d\vec r\cdot\vec A(\vec r)$ is the Peierls phase due to the magnetic
flux $\mathcal{B}_z=\hat{z}\cdot(\vec \nabla\times\vec A)$. We consider
a perturbation $\sum_i\vec B\cdot \vec S_i$ arising from an external Zeeman field $\vec B$, where $\vec S_i=\sum_{ss'}c^\dagger_{is}\vec \sigma_{ss'}c_{is'}$ is the spin at site $i$. Intrinsic electron-electron interactions is furthermore included in the local Hartree-Fock
approximation \cite{Lado:PRB14}, which then take the form of a self-consistent
Zeeman-like field $\vec B_U(\vec r_i)$ that is different in the two honeycomb sublattices. The total Zeeman-like perturbation $H_Z$ thus reads
\begin{equation}\label{meanfield}
H_Z=\sum_{i}\left[\vec B+\vec B_U(\vec r_i)\right]\cdot\vec S_i
\end{equation}
While $\vec B$ is uniform, favoring ferromagnetic (F) ordering, the
self-consistent $\vec B_U(\vec r_i)$ is generally opposite for nearest neighbours, favouring antiferromagnetic ordering of the bulk. The
combination of the two leads to a $\vec B$-tuneable, spin-ordered ZLL that can be tuned from AF to F, as discussed in the main text. \New{Further details on the mean field numerics and results can be found in Ref. \onlinecite{Lado:PRB14}}.

The hybrid graphene/superconductor (SC) system is described by the Hamiltonian
\begin{eqnarray}\label{H}
H&=&H_0+H_Z+\sum_{\vec r_i\in \mathrm{SC}}\left[\Delta_\mathrm{SC}c^\dagger_{i\uparrow}c^\dagger_{i\downarrow}+\mathrm{h.c}\right]
\end{eqnarray}
The Fermi energy in $H_0$ above is $\mu_N$ for $\vec r_i\notin \mathrm{SC}$ in graphene and
$\mu_S$ for $\vec r_i\in \mathrm{SC}$ at the superconductor (also a honeycomb lattice here, although this is not essential, see Appendix \ref{ap:square}). Similarly $B$, $\vec B_U$ and
$\mathcal{B}_z$ are zero in the superconductor. However, gauge invariance demands that  for a finite magnetic flux in graphene, $\Delta_{SC}(\vec r)=\Delta_{SC}\exp\left[-\frac{2e}{\hbar}\int^{\vec r}d\vec r\cdot\vec A(\vec r)\right]$, where $\Delta_{SC}$ is the pairing for zero flux at a given superconductor.

The critical currents \New{for the Fraunhofer patterns} have been calculated in a wide and short graphene Josephson junction, described by a discretized $H$ using an up-scaled $a_0$ for numerical efficiency, following the ideas of Ref. \onlinecite{Liu:PRL15}. The critical
current for each magnetic flux is
calculated as $I_c = 2e/\hbar\times\text{max}_\phi(dF/d\phi)$, where
$\phi$ is the superconducting phase difference and $F(\phi)$ is the free energy of the junction. \cite{Beenakker:92} 
Exact diagonalization of the Hamiltonian is used to evaluate $F(\phi)$ at zero temperature. \New{The method to compute the differential conductance for transport spectroscopy is explained in Appendix \ref{ap:dIdV}}.

\section{Low-energy description of a graphene/superconductor junction}

\label{ap:Heff}

In this Appendix we derive a simplified effective Hamiltonian $H_\mathrm{eff}$ for the two helical edge modes below the superconducting gap $\Delta_\mathrm{SC}$ that arise in a generic Quantum Hall graphene ribbon and a superconductor. We also characterise its topology by deriving expressions for the relevant topological invariant as a function of model parameters.

We assume the chemical potential lies within the gap $\Delta_\mathrm{ZLL}$ induced by interactions in the zero Landau Level (ZLL). The gap is associated to a bulk spin ordering described by a canting angle $\theta$ between the two sublattice. The effective model incorporates an arbitrary value for $\theta$ in graphene and also intervalley coupling due to atomic disorder along the interface. Formally, $H_\mathrm{eff}$ is a projection of the microscopic Hamiltonian on the basis $\{|K\!\!\uparrow\rangle,|K\!\!\downarrow\rangle,|K^*\!\!\uparrow\rangle,|K^*\!\!\downarrow\rangle\}$ of the four AF helical edge states (see Fig. 1f in the main text, and their analytical description in the preceding section) along the junction. We furthermore consider a linearisation of their dispersion in the AF case around `valleys' $K$ and $K^*$ points. These two valleys are  folded onto the $\Gamma$ point by appropriately expanding the ribbon unit cell. Such folding allows us to include intervalley scattering into $H_\mathrm{eff}$ in a simple way. $H_\mathrm{eff}$ then takes the form ($\hbar=1$) 
\begin{equation}\label{Heff}
H_\mathrm{eff}\approx
\left(\begin{array}{cccc}
\mu_1+v_1 k_x & b_\theta & w & 0 \\
b_\theta & \mu_2-v_2 k_x & 0 & w \\
w & 0 & -\mu_2-v_2 k_x & b_\theta \\
0 & w & b_\theta & -\mu_1+v_1 k_x
\end{array}\right)
\end{equation}
Here, the intravalley coupling $b_\theta=\Delta_\mathrm{SC}^*\cos(\theta/2)$ implements AF canting, and couples opposite spins within the same valley.  The intervalley coupling $w$ is spin-independent, and corresponds to the harmonic of wavenumber $\Delta K=(\vec K^*-\vec K)\cdot\hat x$ of any disorder term $W$ close to the interface, $w=\langle\phi_{\uparrow K^*}|W(\Delta K)|\phi_{\uparrow K}\rangle$. Both $b_\theta$ and $w$ can be chosen real without loss of generality. $v_{1,2}>0$ are the velocities of the counter-propagating helical states, and $\mu_{1,2}<\Delta_\mathrm{SC}^*$ are their energy, relative to the Fermi energy, at the $\Gamma$ point (see Fig. 1j in the main text). The overall structure apparent in $H_\mathrm{eff}$ is  fully determined by the particle-hole symmetry of the underlying Nambu description. 
Note that $H_\mathrm{eff}$ only retains terms linear in momentum $k_x$ along the interface.

The simplicity of the linearised low energy model above allows us to compute analytical expressions for the topological invariant and the boundaries that separate different phases of the junction (gapless, trivially gapped, non-trivially gapped), as summarised in the main text. We now briefly sketch their derivation.

\subsection{Topological invariant of edge Hamiltonian}

In symmetry class D (superconductors without time reversal symmetry), the one-dimensional topological invariant is $\mathbb{Z}_2$. It is conventionally defined, for periodic systems, as \cite{Tewari:PRL12}
\begin{equation}\label{Pfaffian}
\nu=\mathrm{sign}\left(\frac{\mathrm{Pf}[H(0)\tau_x]}{\mathrm{Pf}[ H(\pi)\tau_x]}\right)=\frac{s_0}{s_\pi}
\end{equation}
where $s_\alpha=\mathrm{sign}\,\mathrm{Pf}[H(\alpha)\tau_x]$, $H(k_x a_0)$ is the 1D Bloch Hamiltonian for momentum $k_x$, $a_0$ is the lattice constant of the ribbon lattice, $\tau_x$ is the first Pauli matrix in the electron-hole sector, and $\mathrm{Pf}$ is the Pfaffian.  It can be shown in general that $H\tau_x$ is antisymmetric at the high-symmetry points $k_x a_0=0,\pi$. The invariant $\nu$ is thus fully determined by the structure of $H$ at these two points. The effective Hamiltonian $H_\mathrm{eff}$, Eq. (\ref{Heff}), only gives a faithful representation of the full microscopic Hamiltonian $H$ around one of them, the folded $\Gamma$ point $k_x=0$ (Fig. 1j, main text). To extract analytic results for $\nu$ for the full $H$ using $H_\mathrm{eff}$, we must ensure that at the $\pi$-point $s_\pi$ does not change when sweeping the parameter space (since this sector of states is not described by $H_\mathrm{eff}$). This is indeed the case in our system for the chosen basis, for which $s_\pi=1$. The changes in topology stem from the reconnections of the low-energy edge states that are concentrated around $\Gamma$, and are well described by $H_\mathrm{eff}$. Higher excited states not included in $H_\mathrm{eff}$ never cross zero energy, and therefore cannot affect the sign of the Pfaffian of $H(0)\tau_x$. One can thus write
\begin{equation}
\nu=\mathrm{sign}\,\mathrm{Pf}[H_\mathrm{eff}(0)\tau_x]=\mathrm{sign}(w^2+\mu_1\mu_2-b^2)
\end{equation}
We have numerically verified the above result by evaluating the  $\mathbb{Z}_2$ invariant exactly from the microscopic Hamiltonian $H$.

We finally sketch the derivation of the insulating thresholds $w_\mathrm{ins}$ and $\theta_\mathrm{ins}$. These are extracted by computing the solutions for the wavenumber $k_x$ of $H_\mathrm{eff}$ modes at zero energy. Since $H_\mathrm{eff}(k_x)$ is linear in $k_x$, said $k_x$ solutions at $\epsilon=0$ can be obtained as eigenvalues of a matrix $-(\partial_{k_x}H_\mathrm{eff}(0))^{-1}H_\mathrm{eff}(0)$. These can be worked out analytically, and turn out to be all complex (i.e. the interface becomes gapped, see e.g. Fig. \ref{phase}f) if $w>w_\mathrm{ins}$ or $\theta<\theta_\mathrm{ins}$, with the expressions given in the main text,
\begin{eqnarray}
w_\mathrm{ins}&=&\frac{1}{2}\left|\mu_1\sqrt{v_2/v_1}-\mu_2\sqrt{v_1/v_2}\right|\\
b_{\theta_\mathrm{ins}}&=&\Delta_\mathrm{SC}^*\cos(\theta_\mathrm{ins}/2)\equiv\frac{1}{2}\left|\mu_1\sqrt{v_2/v_1}+\mu_2\sqrt{v_1/v_2}\right|\nonumber
\end{eqnarray}

\section{\New{Transport spectroscopy}}
\label{ap:dIdV}

\begin{figure*}
\includegraphics[width=0.65\textwidth]{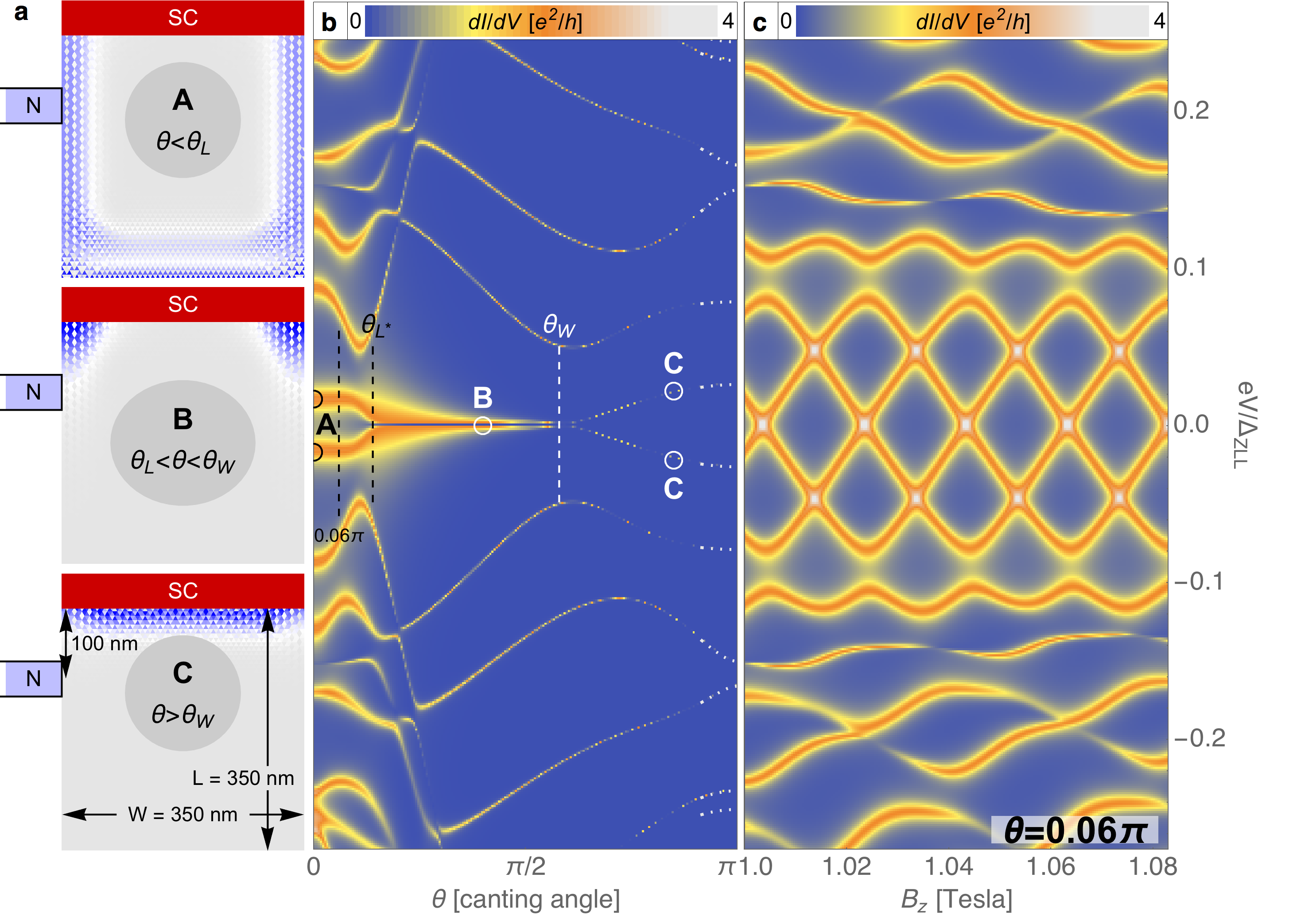}
\caption{\New{(a) 350 nm $\times$ 350 nm sample with a single superconducting contact, otherwise identical to system of Fig. \ref{fig:3}, main text. (b) and (c) show the differential conductance as a function of canting angle $\theta$ and out of plane magnetic field $B_z$. Both show the same phenomenology as panels (b) and (e) of the two contact setup in Fig. \ref{fig:3}, albeit with a simpler two-Majorana geometry, and with the total length of the vacuum edge $L^*=2L+W$ playing the role of $L$.
}}
\label{fig:dIdVap}
\end{figure*}

\subsection{Computing differential conductance}
The computation of the transport spectroscopy results shown in Fig. \ref{fig:3} follows standard techniques of quantum transport. All interactions are incorporated into the mean field solution for $\vec B_U$, Eq. (\ref{meanfield}). In this case, the differential conductance $dI/dV$ through a normal contact with a single spinful channel at a certain bias $V$ is given by the BTK formula $dI/dV=(2-R_{ee}+R_{he})e^2/h$, where the total electron-electron ($R_{ee}$) and electron-hole ($R_{he}$) reflection probabilities of free electrons incident on the contact are evaluated at energy $\epsilon=eV$. The full $R$ matrix, including both electron and hole sectors, may be obtained in terms of Caroli's formula $R=\mathrm{Tr}\left(G_r\Gamma G_a\Gamma\right)$, where $G_{r/a}$ are the (dressed) retarded/advanced Green functions in the sample, and $\Gamma=-i(\Sigma-\Sigma^\dagger)$ is (twice) the decay rate matrix into the normal probe. All these operators are defined in the Nambu basis, just like e.g. Eq. (\ref{hamil}). The Green's function matrices where solved by inverting the Dyson equation $(\epsilon\pm i0^+-H-\Sigma)G_{r/a}(\epsilon)=\mathbb{I}$ using efficient linear algebra routines, where $H$ is the Nambu Hamiltonian of graphene, including the superconductors. The self-energy $\Sigma=V^\dagger g_r V$, defined in terms of the  hopping matrix $V$ between graphene and the semi-infinite normal lead, and the retarded Green function $g_r$ of the latter. $g_r$ is obtained by solving the corresponding (self-consistent) Dyson equation $(\epsilon+i0^+ - h - v^\dagger g_r v)g_r=\mathbb{I}$, where $h$ and $v$ are the on-site and hopping matrices acting on the lead's constituent unit cells.

\begin{figure*}
\includegraphics[width=0.85\textwidth]{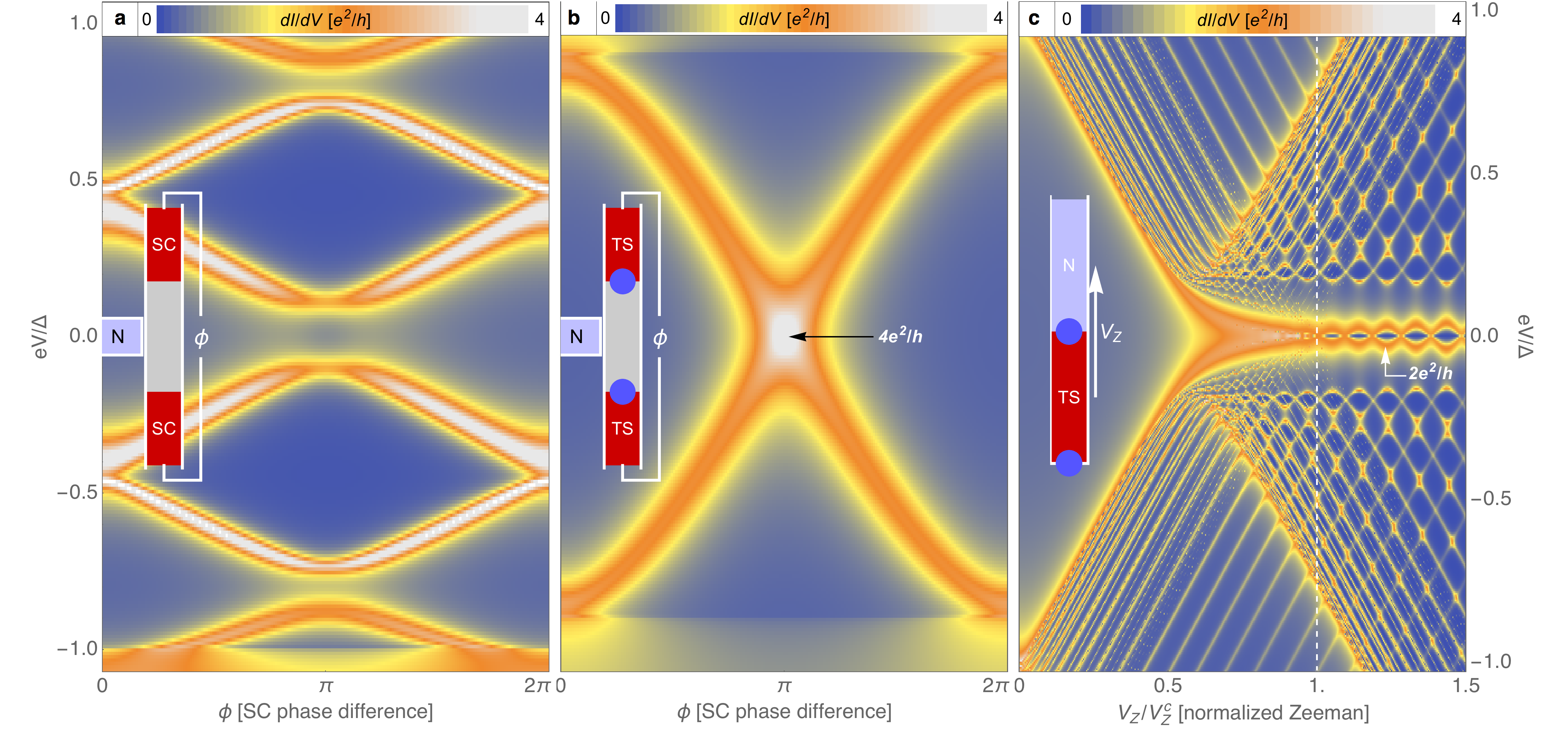}
\caption{\New{Transport spectroscopy $dI/dV$ obtained in various arrangements of normal, trivial superconducting (SC), and topological superconducting (TS) nanowires, to be compared to Fig. \ref{fig:3}. The wires are modelled after Refs. \onlinecite{Oreg:PRL10,Lutchyn:PRL10}. (a) A trivial SC/normal/SC Josephson junction, is probed by a third normal contact at bias $V$, as a function of superconducting phase difference $\phi$. No protected zero-bias $dI/dV$ anomalies aries. (b) Same as (a) for a TS/normal/TS junction. A universal $4e^2/h$ zero bias anomaly (white) is obtained at a certain $\phi$, here $\phi=\pi$, for which the two Majoranas in the junction (blue circles in the inset) decouple. (c) A normal/SC junction that transitions into normal/TS with two Majoranas as the longitudinal Zeeman field $V_Z$ exceeds a critical value $V_Z^c$. Apart from a small oscillatory splitting that decays exponentially with TS length, the $dI/dV$ in the latter case shows a universal $2e^2/h$ zero-bias anomaly (orange).}}
\label{fig:dIdVref}
\end{figure*}

\subsection{Estimates for canting angles $\theta_{L,W}$}

The canting angle $\theta_{L}$ is defined as the $\theta$ such that the corresponding decay length of edge states along a \emph{vacuum} edge equals the length $L$ of the Josephson junction, see Fig. \ref{fig:3}a. Likewise, $\theta_{W}$ is defined as the $\theta$ such that the corresponding decay length of edge states along a \emph{superconducting} edge equals the width $W$ (hence $\theta_W\leq\theta_\mathrm{ins}$).

The evaluation of $\theta_W$ can be made by extracting the decay length $1/\mathrm{Im}k_x$ of gap states at zero energy from the effective Hamiltonian of Eq. (\ref{Heff}) without disorder ($w=0$), and equating that to $W$. The result comes out simply as
\[
\cos\theta_W\approx\cos\theta_\mathrm{ins}+\frac{2v_1v_2}{W^2\Delta_\mathrm{SC}^{*2}}
\] 
where, recall, $\hbar=1$ and
\[
\cos(\theta_\mathrm{ins}/2)=\frac{1}{2}\left|\mu_1\sqrt{v_2/v_1}+\mu_2\sqrt{v_1/v_2}\right|/\Delta_\mathrm{SC}^*
\]
Note that in the limit $W\to\infty$, $\theta_W=\theta_\mathrm{ins}$, as expected.

An analogous calculation can be done for the vacuum edge along the $y$ direction, whose effective (normal) Hamiltonian can be written in analogy to Eq. (\ref{Heff}) as
\begin{equation}
H^\mathrm{vac}_\mathrm{eff}\approx
\left(\begin{array}{cccc}
\mu_N+v_F k_y & \frac{1}{2}\Delta_\mathrm{ZLL}\sin(\theta/2) \\
\frac{1}{2}\Delta_\mathrm{ZLL}\sin(\theta/2) & \mu_N-v_F k_y 
\end{array}\right)
\end{equation}
This leads to the estimate
\[
\sin^2\frac{\theta_L}{2}\approx\left(2\frac{\mu_N}{\Delta_\mathrm{ZLL}}\right)^2+\left(\frac{2v_F}{L\Delta_\mathrm{ZLL}}\right)^2
\]
Note that as $L\to\infty$, $\theta_L$ reaches a minimum value that corresponds to the threshold where the vacuum edge becomes gapped (non-zero for $\mu_N\neq 0$).

Mean field results within the Hubbard model (see Appendix \ref{ap:modelling} and Ref. \onlinecite{Lado:PRB14}) yield a dependence of canting angle $\theta$ with in-plane magnetic field $B_\parallel$ of the form $\sin(\theta/2)\approx \sqrt{1-(B_\parallel/B_0)^2}$, where $B_0$ is the in-plane field that achieves complete Ferromagnetic polarization.

\subsection{Differential conductance with a single superconducting contact}

The formation and detection of graphene-based Majoranas only requires a single superconducting contact. In that sense, the geometry discussed in Fig. \ref{fig:3} of the main text, while relevant in the context of the Josephson effect, Fraunhofer patterns and parity crossings, is not minimal. A simpler geometry with a single superconducting contact allows for the detection of two Majoranas (instead of four) in the form of a zero-bias anomaly, analogous to that of Fig. \ref{fig:3}c. In Fig. \ref{fig:dIdVap} we present the $dI/dV$ in such a geometry. Panel b shows the formation of a zero bias $2e^2/h$ anomaly within $\theta_W<\theta<\theta_{L^*}$ (with a $L^*=2L+W$ that now corresponds to the total length of the vacuum edge).  Note that, while this setup (panel a) does not allow for a $\phi$-controlled modulation, its $\theta$ dependence is qualitative the same as for a Josephson junction. Interestingly, moreover, the analogous to the $\phi$ modulation induced by changing the flux $B_z$ does operate in this setup just like in a Josephson junction. The reason is that the flux changes the relative phase of the two Majoranas in the SC contact, making them cross at zero energy each time the flux $\Phi$ is increased by $\Phi_0$. This is shown in Fig. \ref{fig:dIdVap}c. 

\subsection{Relation of the graphene $dI/dV$ to junctions of topological nanowires}

The transport spectroscopy results for Majoranas in graphene presented in the main text exhibit three different regimes, labeled as $A$, $B$ and $C$ in Figs. \ref{fig:3} and \ref{fig:dIdVap}. These arise due to the interplay, as a function of canting angle $\theta$ between delocalization of Majorana bound states along either a vacuum edge and the superconducting contact. From the point of view of the normal point contact, the former case ($A$) is analogous to a one-dimensional TS/normal/TS Josephson junction, where the normal probe is an extra lead coupled to the normal section for spectrocopy. In contrast, in the case for which the Majoranas remain bound to the corners of the sample and do not delocalize ($B$), the probe is tunnel coupled to the closest Majorana bound state (top-left corner of the sample), and plays the role of a tunnel normal/TS junction like in the zero-bias anomaly experiment of Ref. \cite{Mourik:S12}. Finally, case $C$ is like case $B$ albeit for a trivial normal/S junction without Majorana bound states. These mappings to well-understood systems are useful to understand the universal values of the $dI/dV$ obtained in each case. 

For case analogous to $B$, a one-dimensional normal/TS junction, it is well known \cite{Law:PRL09,Beri:PRB09,Flensberg:PRB10,Fidkowski:PRB12,Pikulin:NJOP12,Pikulin:PRB13,Ioselevich:NJOP13} that the $dI/dV$ for a long enough TS yields a universal $2e^2/h$ zero-bias conductance resonance, a telltale signal of the presence of a Majorana bound state at the contact. If the TS has a finite length, the overlap of the Majorana with its sibling at the opposite end of the TS section gives rise to a splitting of the zero-bias resonance by an energy exponentially small in the TS length divided by the spin-orbit length (or, more precisely, the Majorana localization length\cite{Klinovaja:PRB12}). A simple model for such a N/TS junctions was devised by Oreg \emph{et al.} and by Lutchyn \emph{et al.} in Refs. \onlinecite{Oreg:PRL10,Lutchyn:PRL10}. The model is based on semiconducting  wires that exhibit a TS phase when proximized to an s-wave SC while under a longitudinal Zeeman $V_Z$ exceeding a critical value $V_Z^c=\sqrt{\mu^N+\Delta^2}$. The typical $dI/dV$ in such a junction as a function of $V_Z$ and bias $V$ is shown in Fig. \ref{fig:dIdVref}c. Note that, indeed, for $V_Z>V_Z^c$, a zero bias anomaly of magnitude $2e^2/h$ develops, with a small splitting (oscillating in $V_Z$) \cite{Prada:PRB12, Lim:PRB12,Das-Sarma:PRB12,Rainis:PRB13}.

In the case analogous to $A$, we have a TS/normal/TS, where the normal portion represents the graphene vacuum edge over which the Majoranas delocalize, and probed with an additional point contact. The Oreg-Lutchyn model in the topological phase yield a transport spectroscopy map, shown in Fig. \ref{fig:dIdVref}b, with a single zero energy crossing as the junction phase difference $\phi$ increases by $2\pi$, just like in Fig. \ref{fig:3}(b,d), by virtue of the non-trivial junction topology. Moreover, the zero-bias $dI/dV$ at the crossing is pinned to $4e^2/h$ (white), again like in the graphene case Fig. \ref{fig:3}(b,d). This universal value can be understood intuitively as the addition in parallel of two $2e^2/h$ normal/TS zero bias anomalies, one per Majorana (both are coupled to the probe in this geometry), when the two become decoupled at the appropriate $\phi$. 

Finally, note that the trivial S/normal/S Josephson junction, corresponding to case $C$, does not yield a universal zero-bias anomaly at any phase $\phi$, see Fig. \ref{fig:dIdVref}a. Instead, the finite energy Andreev levels yield a $dI/dV$ at finite bias that approaches (non-universal) $~4e^2/h$. Unlike for the topological case $A$ above, however, perturbations to the system may introduce additional normal-reflection component to said Andreev levels that suppress this value.

\section{Helical edge states in an AF graphene/superconducting interface}
\label{ap:wavefunction}


\subsection{Interface states without Landau levels}

The interface states between a superconductor and an antiferromagnetic honeycomb lattice are not related to the Landau level structure.
In the particular case of graphene, the magnetic field is the key ingredient to develop magnetic order
(due to the large kinetic energy of electrons), which is developed when the kinetic energy
is quenched by the magnetic field.
However, in a general honeycomb lattice, provided the
interactions are large enough
at $B=0$, electrons might be able to develop $AF$ order,
and thus create interface states between a SC even at $B=0$. This
behaviour might be relevant for other honeycomb systems with smaller
hopping strengths than graphene, as silicene, germanene, stanene
or honeycomb oxides.

\begin{figure}
   \centering
   \includegraphics[width=\columnwidth]{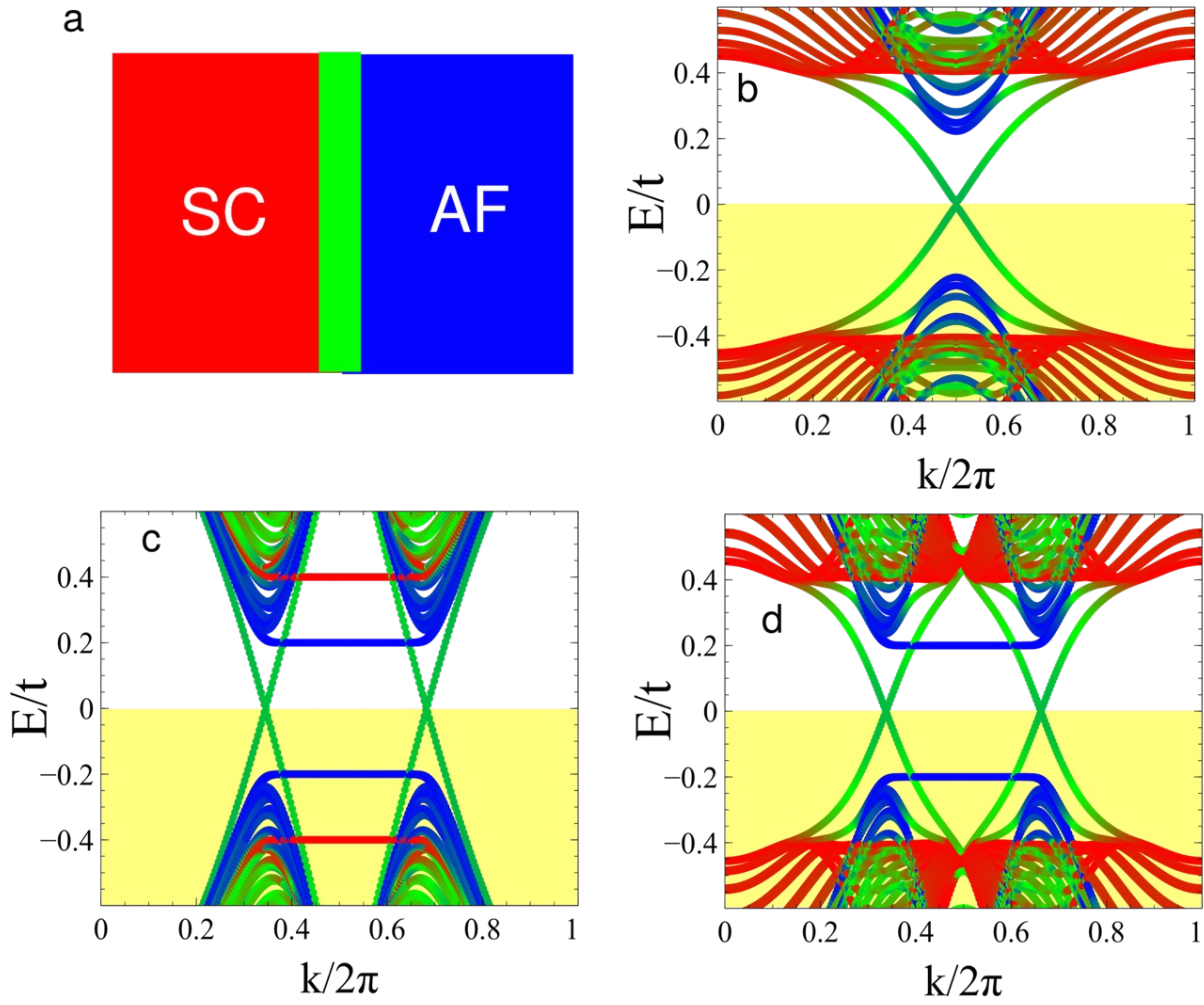}
   \caption{ (a) Scheme of a ribbon hybrid ribbon AF-SC, with the color
code of the bandstructure. (b) Band structure of a hybrid ribbon
with doped SC and armchair interface. Band structure of a hybrid ribbon
with un-doped (c) and doped (d) SC with zigzag interface, which
shows the different interface states in each valley.}
\label{af-sc}
\end{figure}

We show in Fig. \ref{af-sc}, that such interface sustains the same kind of states as in the case
of the antiferromagnetic quantum Hall state. In the case of zigzag interfaces
(\ref{af-sc}c,\ref{af-sc}d), each
valley supports its own set of interface states, whereas for an armchair interface (\ref{af-sc}b),
the two valley are folded and interface states between different valleys can couple.
In the former case, if the interface is abrupt enough, a small gap opens up due to intervalley
mixing. 

If a canting in the magnetic moments is introduced, the zizgag interphase remains gapless. The same
happens when only a orbital magnetic field is introduced. Only when both perturbations
are present simultaneously, the system is able to enter into the
topological superconducting state. Thus, an off-plane magnetic field is mandatory
to observe the Majorana bound states. The previous phenomenology, suggests
that in order to develop a topological gap,
both the spin rotation symmetry and the spatial gauge symmetries have to be broken.

\subsection{Helical edge states from wavematching}
The interface states between a honeycomb antiferromagnet
and a superconductor are not intrinsically related to the
Landau level spectrum. Although in graphene, the antiferromagnetic state is only expected to arise when the system enters in the quantum Hall regime, a general antiferromagnetic
honeycomb lattice might also sustain interface states
when attached to a superconductor without a magnetic flux.

In this section, we will show how that interface states naturally arise
by an analytic argument in the absence of magnetic field. In
particular, a simple wavematching
between a $E=0$ energy state shows that the boundary
between an antiferromagnet and an swave superconductor is able to sustain such
state.

To proceed, we will look for bounded solutions such that $H | \phi \rangle = 0$ with

\begin{equation}
\phi (x) =
 \begin{pmatrix}
 c_1 \\
 c_2 \\
 c_3 \\
 c_4 \\
 \end{pmatrix}
r(x) = 
 \begin{pmatrix}
 c_1 \\
 c_2 \\
 c_3 \\
 c_4 \\
 \end{pmatrix}
 e^{-\lambda(x) x}
\end{equation}

\begin{equation}
\lambda(x) =
  \begin{cases}
    -\Delta_{SC} & \quad \text{if $x<0$}  \\
    m  & \quad \text{if $x>0$} \\
  \end{cases} 
\end{equation}

In the following we will focus in one of the four decoupled sectors,
in particular the $|e,\uparrow K\rangle$ with $|h,\downarrow K'\rangle$
sector

For the antiferromagnet the Hamiltonian reads

\begin{equation}
H_{SC} =
 \begin{pmatrix}
  m & p & 0 & 0 \\
  p & -m & 0 & 0 \\
  0 & 0 & m & -p \\
  0 & 0 & -p & -m \\
 \end{pmatrix}
\end{equation}

whose $E=0$ solutions are

\begin{equation}
\Phi_1 =
\frac{1}{\sqrt{2}}
 \begin{pmatrix}
  1 \\
  i \\
  0 \\
  0 \\
 \end{pmatrix}
\qquad
\Phi_2 =
\frac{1}{\sqrt{2}}
 \begin{pmatrix}
  0 \\
  0 \\
  1 \\
  -i \\
 \end{pmatrix}
\end{equation}

On the other hand, for the superconductor the Hamiltonian reads

\begin{equation}
H_{SC} =
 \begin{pmatrix}
  0 & p & \Delta_{SC} & 0 \\
  p & 0 & 0 & \Delta_{SC} \\
  \Delta_{SC} & 0 & 0 & -p \\
  0 & \Delta_{SC} & -p & 0 \\
 \end{pmatrix}
\label{hamil}
\end{equation}

with $E=0$ solutions

\begin{equation}
\Psi_1 =
\frac{1}{\sqrt{2}}
 \begin{pmatrix}
  0 \\
  1 \\
  i \\
  0 \\
 \end{pmatrix}
\qquad
\Psi_2 =
\frac{1}{\sqrt{2}}
 \begin{pmatrix}
  1 \\
  0 \\
  0 \\
  i \\
 \end{pmatrix}
\end{equation}

Imposing continuity at the interface $x=0$, the full $E=0$ solution reads

\begin{equation}
 \phi (x) =
\frac{1}{2} 
 \begin{pmatrix}
  1 \\
  i \\
  -1 \\
  i \\
 \end{pmatrix}
r (x)
\end{equation}

\begin{equation}
r (x) =
  \begin{cases}
    e^{\Delta_{SC} x} & \quad \text{if $x<0$}  \\
    e^{-m x}  & \quad \text{if $x>0$} \\
  \end{cases} 
\end{equation}

so that a normalizable $E=0$ exist for an interface between a trivial antiferromagnet and a trivial Dirac superconductor.

\subsection{Helical modes from explicit integration}
In previous section, we build the $E=0$ by wavematching across a sharp interface.
However, it is is possible to give a general solution for the interface state
between the antiferromagnet and the superconductor. Without
loss of generality, in the following
we will assume $m>0$ and $\Delta_{SC}>0$. The Hamiltonian
for an arbitrary antiferromagnet and pairing profile for $k_y=0$ reads

\begin{equation}
H = \gamma_1 m (x) + \gamma_2 p + \gamma_3 \Delta_{SC} (x) 
\end{equation}

with $\gamma_{1},\gamma_2,\gamma_3$ defined by 

\begin{equation}
\gamma_1 =
 \begin{pmatrix}
  1 & 0 & 0 & 0 \\
  0 & -1 & 0 & 0 \\
  0 & 0 & 1 & 0 \\
  0 & 0 & 0 & -1 \\
 \end{pmatrix}
\end{equation}

\begin{equation}
\gamma_2 =
 \begin{pmatrix}
  0 & 1 & 0 & 0 \\
  1 & 0 & 0 & 0 \\
  0 & 0 & 0 & -1 \\
  0 & 0 & -1 & 0 \\
 \end{pmatrix}
\end{equation}

\begin{equation}
\gamma_3 =
 \begin{pmatrix}
  0 & 0 & 1 & 0 \\
  0 & 0 & 0 & 1 \\
  1 & 0 & 0 & 0 \\
  0 & 1 & 0 & 0 \\
 \end{pmatrix}
\end{equation}

Defining $\gamma_4 = -i \gamma_2\gamma_1$ and
$\gamma_5 = i \gamma_2\gamma_3$, the zero energy equation reads

\begin{equation}
(p + i\gamma_4 m (x) - i\gamma_5 \Delta_{SC} (x)) \phi = 0
\end{equation}

The spinor wavefunction

\begin{equation}
 \phi_0 =
\frac{1}{2} 
 \begin{pmatrix}
  1 \\
  i \\
  -1 \\
  i \\
 \end{pmatrix}
\end{equation}

verifies $\gamma_4 \phi_0 = \gamma_5 \phi_0 = \phi_0$, which
allows to build the $E=0$ solution

\begin{equation}
\phi (x) = N e^{-\int_0^x ( m(x') - \Delta_{SC} (x') ) dx'} \phi_0
\end{equation}

which is normalizable provided that $m(+\infty) > \Delta_{SC} (+\infty)$
and $\Delta_{SC} (-\infty) > m(-\infty)$, which is the condition
of domain wall between superconductor and antiferromagnet. For the
case of step profiles, the solution obtained by wavematching is recovered.


\begin{figure}
   \centering
   \includegraphics[width=\columnwidth]{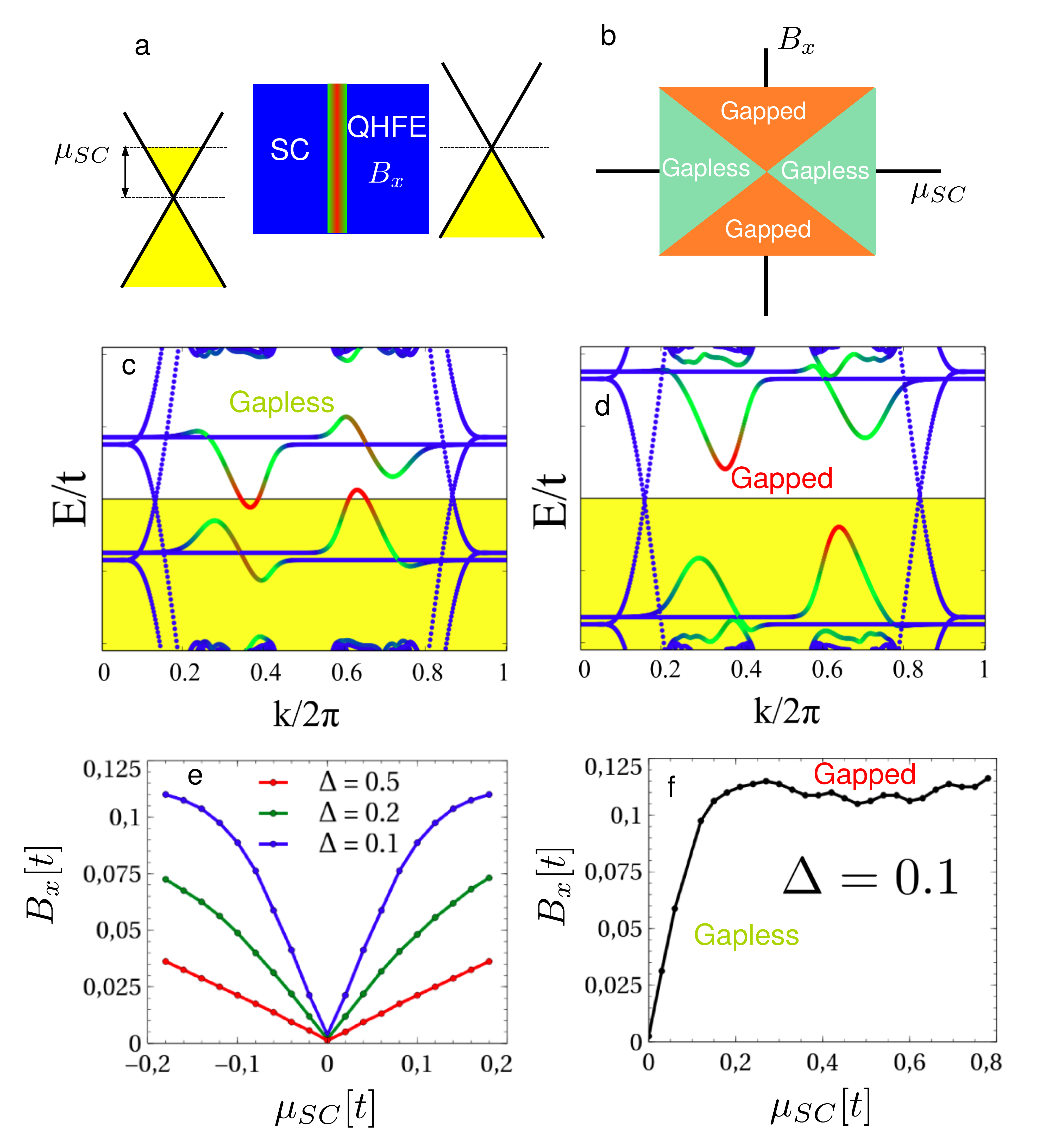} 
   \caption{Scheme of the effect of chemical doping in the Dirac spectrum
of proximized graphene (a). Schematic phase diagram (b) of the interface
electronic spectrum 
between the SC and the QH ferromagnet,
as function of the in-plane field and the chemical
doping of the SC. 
Quasiparticle energies
for a gapless (c) and gapped (d) interface. 
Phase boundary obtained by numerical calculation
at low dopings (e) and at large doping (f). 
The green and red colored states
of the band structures
correspond to the states localized at the interface, where the topological superconducting gap
is calculated between the red states shown in (c,d). The blue gapless states
correspond to the chiral states between the QH and vacuum.}
\label{phase}
\end{figure}

\subsection{Influence of $\mu_{SC}$ in the critical Zeeman}

In the idealised situation in which the superconductor is described as a single-orbital honeycomb
lattice at half filling, a arbitrary small Zeeman field is capable of
opening the interface topological gap. However, charge transfer processes
are expected to shift the chemical potential of the proximized graphene \New{(SC region in Fig. \ref{fig:1})}.
In this situation, band bending of the interfacial states
leads
to a one dimensional gapless state (see Fig. \ref{phase}c).
In order to reach the interfacial topological superconducting state,
the bended bands (Fig. \ref{phase}c) have to be moved up in energy.
This can achieved by increasing the in-plane field, \New{so that $\theta<\theta_\mathrm{ins}$} .
The critical in-plane field $B_x$ as a function of doping,
which separates the gapless and topologically gapped states
is shown in Figs. \ref{phase}e,f.
For small $\mu_{SC}$, the critical field increases linearly,
leading to small critical fields at small doping in the SC, whereas
for large doping the critical field saturates.

\begin{figure}
   \centering
   \includegraphics[width=\columnwidth]{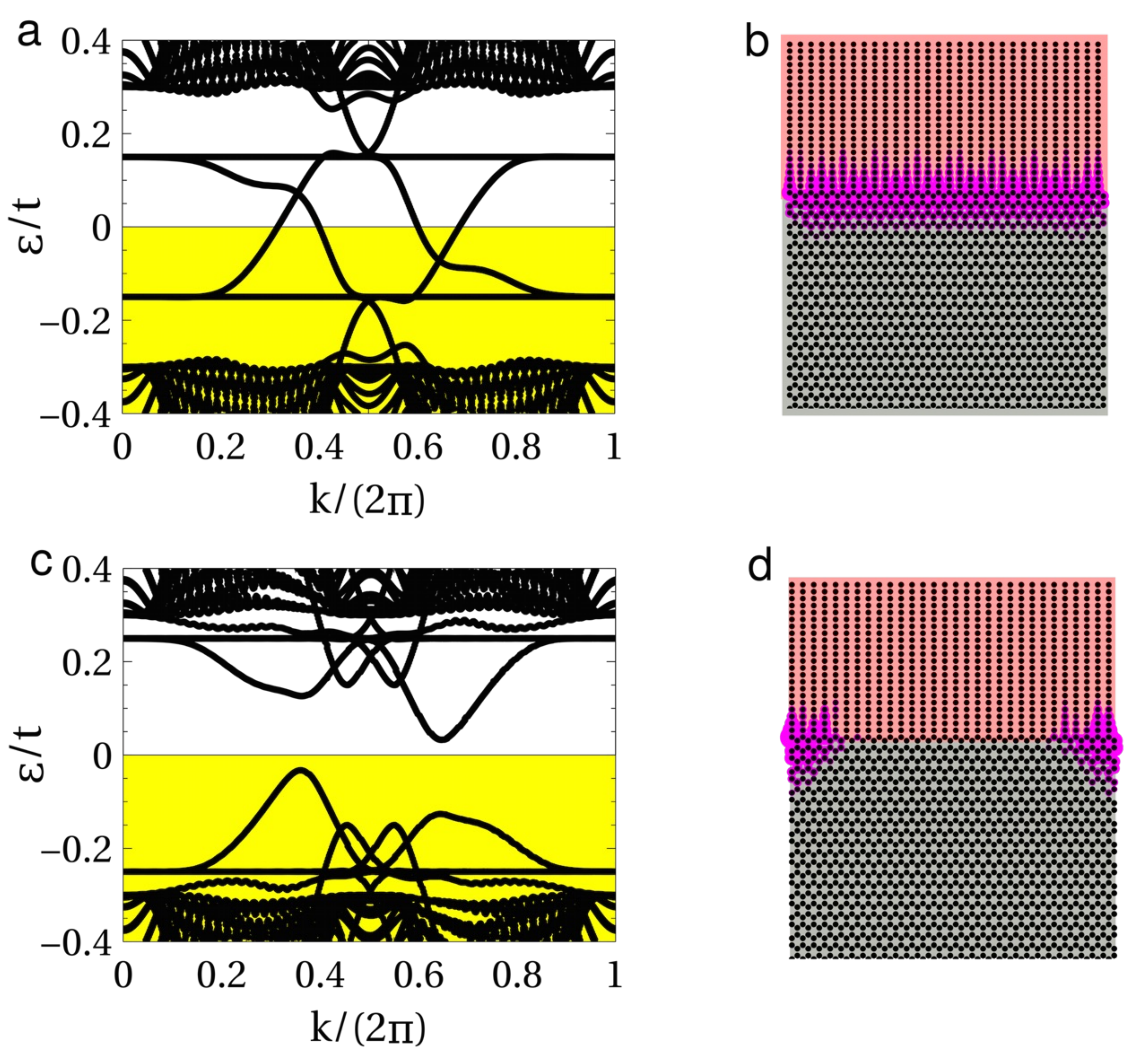} 
   \caption{Bandstructure of the an interface between a square superconductor
and an antiferromagnet (a) or canted antiferromagnet (c) honeycomb lattice.
Figures (b) and (d) show the local density of states in a finite system,
which correspond to gapless interface channels (without canting) (b) and Majorana states (with canting) (d) } 
\label{square}
\end{figure}

\subsection{Emergence of AF helical states from a topological point of view}

\New{An analysis of the emergent AF helical edge modes in terms of topology can be made, but it is less rigorous mathematically than the topological superconductor order in the canted AF phase. As shown in Fig. \ref{fig:1}k, there is a finite volume in parameter space for which the SC contacts are helical metals. The relevant symmetry class within the ten-fold way\cite{Altland:PRB97} for the infinite system is the 2D class D, and its topological invariant is $\mathbb{Z}$, which corresponds to the number of \emph{chiral} edge states at a surface \cite{Schnyder:PRB08}. Within this language the helical SC contact has a trivial (zero) $\mathbb{Z}$ invariant, since the number of right minus left propagating modes is zero. This is actually the reason why crossing the $\theta=\theta_\mathrm{ins}$ destroys the helical edge states without an intervening bulk-gap inversion. Therefore, the reason for the existence of helical states for $\theta>\theta_\mathrm{ins}$ cannot be found in the standard homotopy classification. It is rather an instance of non-trivial \emph{valley} Chern number.}

\New{If one computes the $\mathbb{Z}$ invariant in 2D of our system, both in the graphene side and on the superconductor side, one needs to integrate the Berry curvature of the Nambu bands. For the superconductor one obtains negligible Berry curvature for all momenta. However on the graphene side (and choosing the magnetic unit cell to compute the bands), one finds that while the integrated curvature is zero (hence $\mathbb{Z}$ is zero), it is the sum of two integer and opposite contributions from different valleys. For a specific spin sector (e.g. $|e\uparrow\rangle, |h\downarrow\rangle$), one valley has partial integral 1 and the other -1. Therefore, assuming valley symmetry is preserved at the graphene/SC interface (that is $w=0$ in the phase diagram of Fig. \ref{fig:1}k, i.e. the contact is transparent), one can invoke a bulk-boundary correspondence principle for each of the two valleys and spin sectors independently, which yields two pairs of counterpropagating (helical) states (one per valley and spin). If the contact is not transparent ($w>0$) but intervalley scattering is below the threshold $w<w_\mathrm{ins}$, the helical states will split, but the contact will still be metallic (since the splitting is smaller than the energy where they cross). Hence the AF helical states are an instance of weak topology from the two valleys, just like e.g. the helical modes in works like Refs. \cite{Martin:PRL08, San-Jose:PRB13}. Note, however, that the mathematical standing of these arguments are less sound than the conventional ones from full-Brillouin-zone invariants, since to our knowledge there is no rigorous theorem that guarantees the existence of surface states from partially integrated (valley) Chern numbers.}

\section{Square-lattice superconductor}
\label{ap:square}

In Appendix \ref{ap:wavefunction} we have considered the superconductor arising
from electrons hopping in a honeycomb lattice
and subjected to  pairing potential. Nevertheless,
the fact that the interface states persist
even upon doping of the superconductor, suggests that
their existence goes far beyond
what our analytic argument might suggest. 
Actually, we here show that a honeycomb superconducting
lattice is not mandatory,
so that even an interface between a square superconductor will give rise
to localized Majorana states.

To illustrate this, we show in Fig. \ref{square} the bandstructure
of an interface between the canted Quantum Hall antiferromagnet, and
the local density of states for a finite system. In the case of fully collinear
antiferromagnetism, the interface sustains a gapless channel. When the moments
are canted, a topological gap in the inter-facial bands opens up
and localized Majorana modes show up. 

\bibliography{biblio}

\end{document}